\documentclass[reqno]{amsart}

\usepackage{a4,amsmath,amssymb,amsthm}
\usepackage{xfrac}

\usepackage{amsaddr}

\usepackage[T1]{fontenc}
\usepackage[section]{placeins}
\usepackage{tikz}
\usetikzlibrary{shapes,decorations}
\usetikzlibrary{cd}
\usetikzlibrary{arrows}

\usepackage{ulem}
\usepackage{multicol}

\makeatletter
  \def\moverlay{\mathpalette\mov@rlay}
  \def\mov@rlay#1#2{\leavevmode\vtop{%
     \baselineskip\z@skip \lineskiplimit-\maxdimen
     \ialign{\hfil$#1##$\hfil\cr#2\crcr}}}
\makeatother

\usepackage{stmaryrd}

\usepackage{url}
\usepackage{hyperref}

\usepackage{color}

\numberwithin{equation}{section}

\usepackage[mathscr]{eucal}

\usepackage{graphicx}
\graphicspath{pic/}
\usepackage{psfrag}
\usepackage{blindtext}
\usepackage[inline]{enumitem}
\usepackage[labelfont=bf]{subfig}

\newcommand{\Reals}{{\mathbb R}}         %the real numbers
\newcommand{\scri}{\mathcal{I}}

\theoremstyle{plain}
\newcounter{mnotecount}[section]

\newcounter{mymnotecount}[section]

\title{The \texorpdfstring{$\Lambda$}{Lambda} to Zero Limit of Spacetimes and its Physical interpretation}
\author[M. Bugden, C.~F. Paganini]{Mark Bugden$^\dagger$ and Claudio F. Paganini$^\ddagger$} 
\email{mark.bugden@anu.edu.au}
\email{claudio.paganini@aei.mpg.de}

\address{$^\dagger$Mathematical Sciences Institute, Australian National University, Canberra ACT 0200, Australia }
\address{$^\ddagger$Albert Einstein Institute, Am M\"uhlenberg 1, D-14476 Potsdam,  Germany }

\begin{document}

\date{\today \ {\em File:\jobname{.tex}}}

\begin{abstract} We study the $\Lambda \to 0$ behaviour of Schwarzschild-de Sitter spacetime and show, according to Geroch's notion of spacetime limits, that it converges to the Schwarzschild spacetime. We use an embedding into $AdS_3$ to illustrate and quantify this limiting behaviour. We use these quantitative observations to establish a hierarchy of validity between the Einstein-de Sitter equations and the Einstein equations (and therefore in a weak field limit also Newton's equations), analogous to the quantum-classical limit when taking $\hbar \to 0$. 
\end{abstract}

\maketitle

\tableofcontents

%according to current paradigm we live in a de sitter universe, dark energy is compatible with a cosmological constant.

\section{Introduction}
The currently accepted paradigm in observational astronomy is that the universe in which we live is undergoing an accelerated expansion. Recent data for the Hubble constant from CMB data \cite{planck2016astronomy} and for the Hubble constant from local data \cite{riess20162} are in support of this. The simplest theoretical model incorporating such an accelerating universe is the $\Lambda$CDM model, see e.g. \cite{buchert2016observational} for a review and open tensions. Despite the fact that there exist alternative explanations for the accelerated expansion, such as \cite{racz2017concordance}, we will adopt in the present paper the view that $\Lambda$CDM is a correct description of the expansion of the universe, and that the Einstein-de Sitter equations are the fundamental equations describing gravity.
In the present paper we will argue that the cosmological constant can be assigned a similar function for the gravitational realms that $\hbar$ plays for matter. The correspondence principle in quantum mechanics is the notion that when the scales of the action in a quantum mechanical system become large compared to $\hbar$, the system approximates a corresponding classical system. This quantum-classical correspondence gives a heuristic for recovering a classical system from a quantum system - simply take the limit $\hbar \to 0$. In the present paper we will study the limits of Schwarzschild-de Sitter black holes as the cosmological constant goes to zero, and we will argue that this allows one to define a scale of validity for the Einstein equations of an isolated gravitational system in a de Sitter universe. To do so, we will employ Geroch's notion for the limits of a family of spacetimes \cite{geroch_limits_1969} applied to Schwarzschild-de Sitter. 
To study the way in which the limit is approached in detail, we use an embedding of the quotient of Schwarzschild-de Sitter space over the sphere into $AdS_3$ space.\footnote{That is, $(2+1)$-dimensional anti-de Sitter space.} This embedding was first introduced in \cite{bengtsson_classics_2014} for the case of Reissner-Nordstr\"{o}m.
For calculations on the scale of astrophysical systems the cosmological constant is usually dropped and the spacetime is assumed to be asymptotically flat. This approximation is often employed with little justification, other than a brief citation of the small value $\Lambda \sim 10^{-52} \, m^{-2}$ for the cosmological constant. Recent work that takes the cosmological setting into consideration suggests that the effects are by no means trivial. Prominent examples being the quadropole formula for gravitational energy loss \cite{ashtekar2014asymptotics,ashtekar2015asymptotics,ashtekar2015asymptotics2,ashtekar2016gravitational}, as well as recent work on the gravitational memory effect in de Sitter spacetimes \cite{bieri2016gravitational}. Note that even when one assumes that the universe is spatially flat, asymptotic flatness has to be employed with care, since the definition of asymptotic flatness incudes the requirement that the matter density falls off sufficiently fast towards infinity. This condition is obviously violated for a spacelike slice in a homogeneous, spatially flat FLRW universe.\footnote{This was pointed out to the authors by Beatrice Bonga in private communication} Interestingly, the problem of global non-linear stability for black holes has recently been solved for slow-rotating black holes in a de Sitter universe \cite{hintz2016global},\footnote{This is arguably the physically relevant case if the cosmological constant is, in fact, positive.} while the corresponding problem for asymptotically flat spacetimes remains one of the big challenges in the field of mathematical relativity, see \cite{ma2017uniform,ma2017uniform2,aksteiner2016new,andersson2017morawetz,dafermos2017boundedness,finster2016linear} for recent progress on the linearised problem and \cite{klainerman2017global} for the full non-linear problem under strong constraints. In this paper we will use the qualitative properties of how the $\Lambda \to 0$ limit is approached to give a heuristic argument that the Einstein equations are a legitimate approximation to the fundamental Einstein-de Sitter equations, for calculations in the short-range regime. For gravitational memory this was recently worked out in \cite{bieri2017gravitational}, where the authors found that for low redshift, i.e. for nearby sources, and high frequencies the gravitational memory in a $\Lambda$CDM background is equivalent to that in a flat space while for large redshift there is a significant deviation. 

\subsection*{Overview of the paper}
The paper is organized in the following way. In Section \ref{sec:ssds} we will introduce and review the relevant background, including the Schwarzschild-de Sitter spacetimes. Then, in Section \ref{sec:math}, we discuss Geroch's notion for the limits of spacetimes. In Section \ref{sec:embedding} we discuss how the embedding of Schwarzschild-de Sitter into $AdS_3$ is performed. The resulting embeddings are then presented in Section \ref{sec:pics}. Finally in Section \ref{sec:phys}, we give a possible physical interpretation of our findings. 

%%%%%%%%%%%%%%%%%%%%%%%%%%%%%%%%%%%%%%%%%%%%%%%%%%%%%%%%%%%%%%%%%%%%%%%%%%%%%%%%%%%%%%%%%%%%%%%%%%%%%%%%%%%%%%%%%%%%%%%%%%

\section{The Schwarzschild-de Sitter Spacetime} 
\label{sec:ssds}
\begin{figure}
    \centering
    \includegraphics[width=0.8\textwidth]{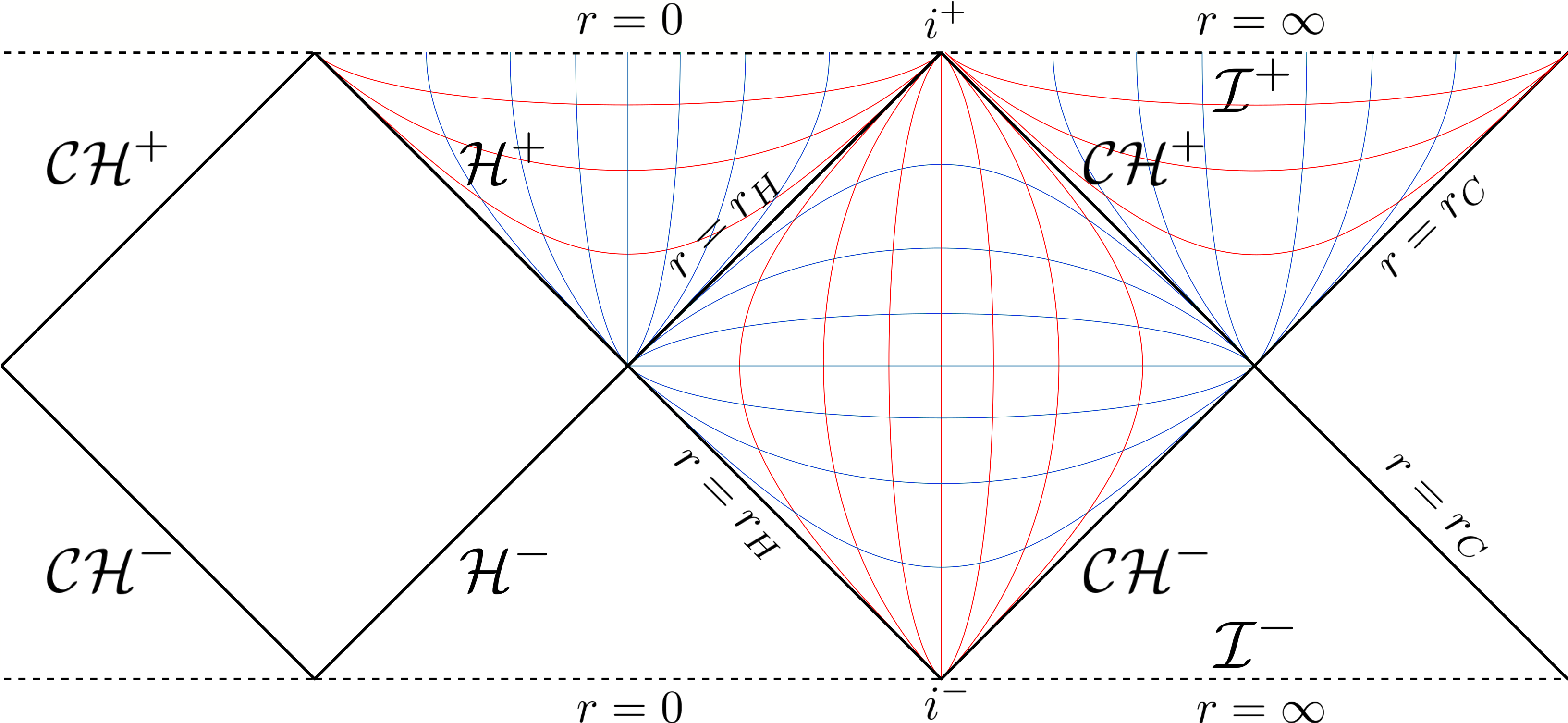}
    \caption{Conformal diagram for the maximal extension of the subextremal Schwarzschild-de-Sitter space-time. The blue lines correspond to hypersurfaces of constant $t$ the red lines to hypersurfaces of constant $r$. $\mathcal{H^\pm}$ are the future and past event horizon located at $r=r_H$ while $\mathcal{CH}^\pm$ are the future/past cosmological horizons located at $r=r_C$. Time like future and past infinity is indicated by $i^\pm$. The singularity is located at $r=0$. Here $r=\infty$ is a spacelike conformal boundary }
    \label{fig:ssds}
\end{figure}

The Schwarzschild-de Sitter spacetime is the spherically symmetric solution to the vacuum Einstein-de Sitter equations\footnote{Note that, until section \ref{sec:phys}, we will use units such that $\hbar = G = c = 1$.}
\begin{equation}
R_{\mu \nu} - \frac{1}{2}R g_{\mu \nu} + \Lambda g_{\mu \nu} = 0
\end{equation}
with $\Lambda >0$. In Schwarzschild coordinates the metric is given by
\begin{equation}\label{eq:ssds}
ds^2 = -f(r) dt^2 + \frac{1}{f(r)} dr^2 + r^2 d \Omega^2
\end{equation}
with
\begin{equation}
f(r) = 1-\frac{2M}{r} - \frac{\Lambda}{3}r^2,
\end{equation}
where $M$ and $\Lambda$ are regarded as free parameters. The spacetime is spherically symmetric and static. We define the domain of outer communication as the region where the Killing vector field $\partial_t$, for which the orbits of points under the diffeomorphism are open, is timelike. The metric \eqref{eq:ssds} has a coordinate singularity when
\begin{equation} \label{eq:radialfunc}
1-\frac{2M}{r} - \frac{\Lambda}{3} r^2 = 0,
\end{equation}
where the norm of the Killing vector $\partial_t$ switches sign, indicating the location of a horizon. Note that this equation always has at least one real solution independent of the choice of parameters. For parameters in the subextremal range there are three real solutions to equation \eqref{eq:radialfunc}. They can be written explicitly as 
\begin{align}
\label{eq:rH}
r_H &= \frac{2}{\sqrt{\Lambda}} \cos \left[\frac{1}{3} \arccos (3M\sqrt{\Lambda}) + \frac{\pi}{3} \right] \\
r_C &= \frac{2}{\sqrt{\Lambda}} \cos \left[\frac{1}{3} \arccos (3M\sqrt{\Lambda}) - \frac{\pi}{3} \right] \\
r_U &= -(r_H + r_C).
\end{align}
In this work we are only interested in the coordinate range where $r\in(0,\infty)$ and, since $r_U$ is always negative, it will not be relevant to our discussion.
In the subextremal case, $r_H$ is the location of the black hole horizon and $r_C$ is the location of the cosmological horizon. It is the region between those two where the Killing vector field $\partial_t$ is timelike. Note that Schwarzschild-de Sitter becomes extremal when $r_H$ and $r_C$ coincide, which is the case when $9\Lambda M^2=1$. We will primarily restrict ourselves to the subextremal case, where $0< \Lambda < \frac{1}{9M^2}$. Note that the photon sphere in Schwarzschild-de Sitter is located at 
\begin{equation}
    r_{ph}=3M,
\end{equation}
independent of the value of $\Lambda$.\footnote{See \cite{geometryphotonsurfaces} for a derivation.} 
The conformal diagram for Schwarzschild-de Sitter is given in Figure \ref{fig:ssds} from which we can see immediately, by gluing two consecutive cosmological horizons together, that its topology is given by $S^1\times S^2\times \Reals$.

\begin{figure}[h!]
    \centering
    \includegraphics[width=0.6\textwidth]{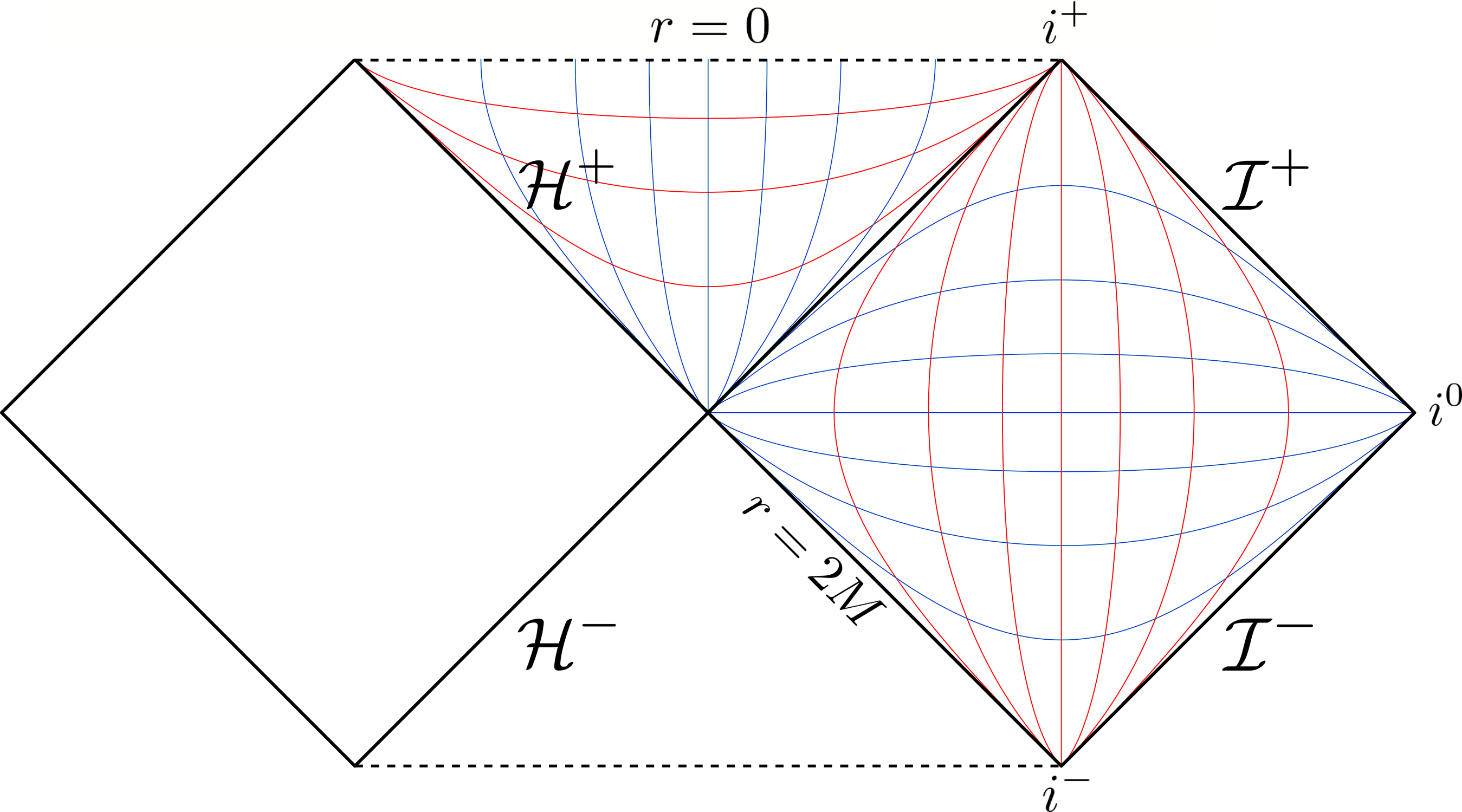}
    \caption{Conformal diagram for the maximal extension of the Schwarzschild space-time. The blue lines correspond to hypersurfaces of constant $t$ the red lines to hypersurfaces of constant $r$. $\mathcal{H^\pm}$ are the future and past event horizon located at $r=2M$ while $\scri^\pm$ are the future/past null infinity. Time like future and past infinity is indicated by $i^\pm$, while $i^0$ indicates space like infinity. The singularity is located at $r=0$.}
    \label{fig:ss}
\end{figure}

In the following when we speak about limits of spacetime properties, we are simply discussing the properties of coordinate functions. This is not to be confused with the limits of spacetimes that we consider later on in the paper, although some intuitive results do carry over. 
Since the location of the photon sphere is constant for a fixed $M$, it is not surprising that in the limit $\Lambda \to \frac{1}{9M^2}$, the two relevant horizons approach this value:
\begin{align*}
\lim_{\Lambda \to \frac{1}{9M^2}} r_H  &= 3M  \\
\lim_{\Lambda \to \frac{1}{9M^2}} r_C &= 3M.
\end{align*}
On the other hand, the limit $\Lambda \to 0$ for these functions is
\begin{align*}
\lim_{\Lambda \to 0} r_H  &= 2M  \\
\lim_{\Lambda \to 0} r_C &= \infty.
\end{align*}
In this limit, the radius of the black hole horizon takes the same value as the black hole horizon from the Schwarzschild metric, for which the function $f(r)$ in the metric \eqref{eq:ssds} is given by
\begin{equation} 
f(r)=1-\frac{2M}{r}.
\end{equation}
The domain of outer communication for Schwarzschild stretches out an infinite distance from the black hole horizon, consistent with the cosmological horizon extending to infinity. The Schwarzschild metric solves the Einstein vacuum equations
\begin{equation}
R_{\mu \nu} = 0,
\end{equation}
and is asymptotically flat. Its conformal diagram is given in Figure \ref{fig:ss}.

%%%%%%%%%%%%%%%%%%%%%%%%%%%%%%%%%%%%%%%%%%%%%%%%%%%%%%%%%%%%%%%%%%%%%%%%%%%%%%%%%%%%%%%%%%%%%%%%%%%%%%%%%%%%%%%%%%%%%%%%

\section{Limits of spacetimes} 
\label{sec:math}
Lorentzian metrics appearing in general relativity often come in families parameterised by one or more constants, whose values are not fixed by the Einstein field equations. Consider, for example, the Kerr family of solutions. In this family, there are two free parameters, corresponding to the mass $M$ and the rotation parameter $a$. It is a natural question to ask what type of spacetime we obtain if we reduce, say, the rotation parameter $a$ to 0. 

Na\"{i}vely, the answer to this question consists of simply setting $a = 0$ in the coordinate description of the metric.\footnote{Or taking the limit $a \to 0$ if required.} This approach has significant issues however, since one can first perform a coordinate transformation and then take the same limit to obtain a completely different spacetime! This fact seems at odds with the notion that coordinate changes in general relativity aren't supposed to affect anything.

Geroch provides the resolution to this paradox by asserting that it is only meaningful to take limits if we first introduce a method of comparing points in different spacetimes \cite{geroch_limits_1969}. That is, we need a way of deciding which points are `the same' in spacetimes which have different values for the chosen parameter. There is no canonical way of doing this, and so any such limit will implicitly involve a choice. 

\begin{figure}[h!]
    \centering
    \includegraphics[width=\textwidth]{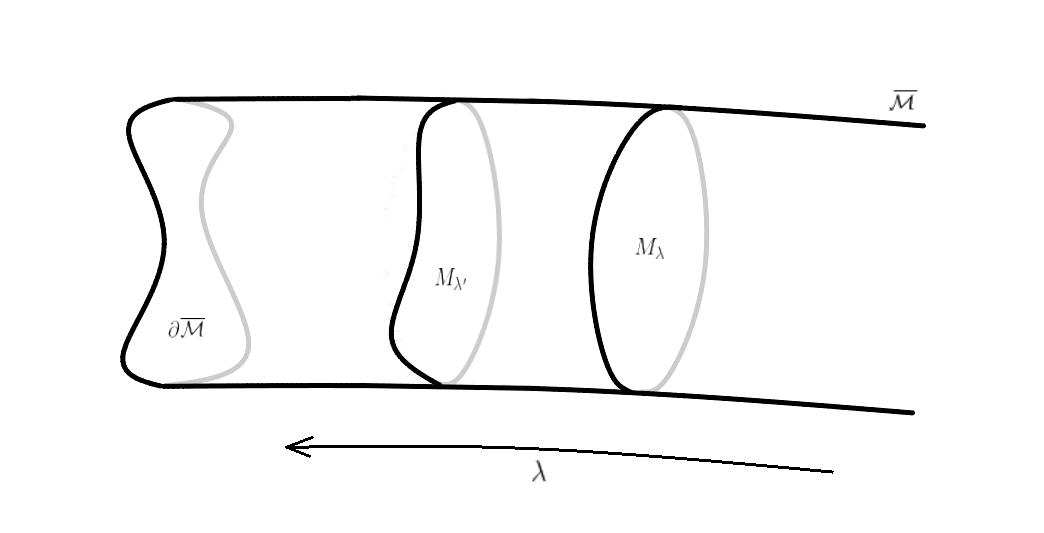}
    \caption{A cartoon depiction of the Geroch foliation.}
    \label{fig:foliation}
\end{figure}

Let us now describe Geroch's prescription in a little detail. We begin with a one-parameter family of spacetimes $M_{\lambda}$, and wish to assign a sensible limiting spacetime to this family as we take the parameter to some fixed value, say $\lambda \to 0$. We assemble the family of spacetimes into a smooth 5-dimensional manifold, $\mathcal{M}$, where each $M_{\lambda}$ is a smooth 4-dimensional submanifold of $\mathcal{M}$.\footnote{Note that unless otherwise specified, we assume all manifolds are Hausdorff.} The manifold $\mathcal{M}$ is foliated by these submanifolds, and the parameter $\lambda$ defines a scalar field on $\mathcal{M}$ which is constant on each leaf of the foliation. We assume the metric tensors $g_{ab} (\lambda)$ combine to form a smooth metric $\mathcal{G}$ on $\mathcal{M}$ with signature $(0, -, +, +, +)$. The data defined by $(\mathcal{M}, \mathcal{G})$ is equivalent to the data defined by the family $(M_{\lambda}, g(\lambda))$. A limiting spacetime is then obtained by defining a suitable boundary $\partial \mathcal{M}$ for $\mathcal{M}$, see Figure \ref{fig:foliation}. More specifically, a limit space is a 5-dimensional manifold $\overline{\mathcal{M}}$ with boundary $\partial \overline{\mathcal{M}}$, a metric $\overline{\mathcal{G}}$ and a scalar field $\overline{\lambda}$ on $\overline{\mathcal{M}}$, and a smooth injective map $\Psi$ from  $\mathcal{M}$ into the interior of  $\overline{\mathcal{M}}$ satisfying:
\begin{itemize}
    \item $\Psi$ takes $\mathcal{G}$ into $\overline{\mathcal{G}}$, and $\lambda$ into $\overline{\lambda}$ \\
    \item $\partial \overline{\mathcal{M}}$ is connected, non-empty, and $\overline{\lambda} = 0$ when restricted to $\partial \overline{\mathcal{M}}$. \\
    \item $\overline{\mathcal{G}}$ has signature $(0,-,+,+,+)$ on $\partial \overline{\mathcal{M}}$.
\end{itemize}

Geroch goes on to define a \emph{family of frames} - that is, for each leaf of the foliation, one chooses a fiducial point $p_{\lambda}$ and an orthonormal frame $\omega(\lambda)$ at $p_{\lambda}$, and identifies such points and frames for each $\lambda$. Then, by calculating geodesics from the fiducial point to any other point, we have a way of comparing points in the different spacetimes. Geroch then states that such a choice of a family of frames either defines no limit space, or else determines a unique maximal limit space. 

How does this connect to our intuitive notion of simply taking the limits in the coordinate representation of the metric? Choosing a coordinate system is implicitly choosing a point,\footnote{Actually, it implicitly chooses any one of the points in the open set on which the coordinates are defined.} and an orthonormal frame at that point, for each value of the parameter $\lambda$. Such a choice of coordinates therefore determines a family of frames, and by Geroch's theorem, a limiting spacetime. There is no guarantee that a different choice of coordinates will result in the same limiting spacetime. 

\begin{figure}[h!]
    \centering
    \includegraphics[width=0.6\textwidth]{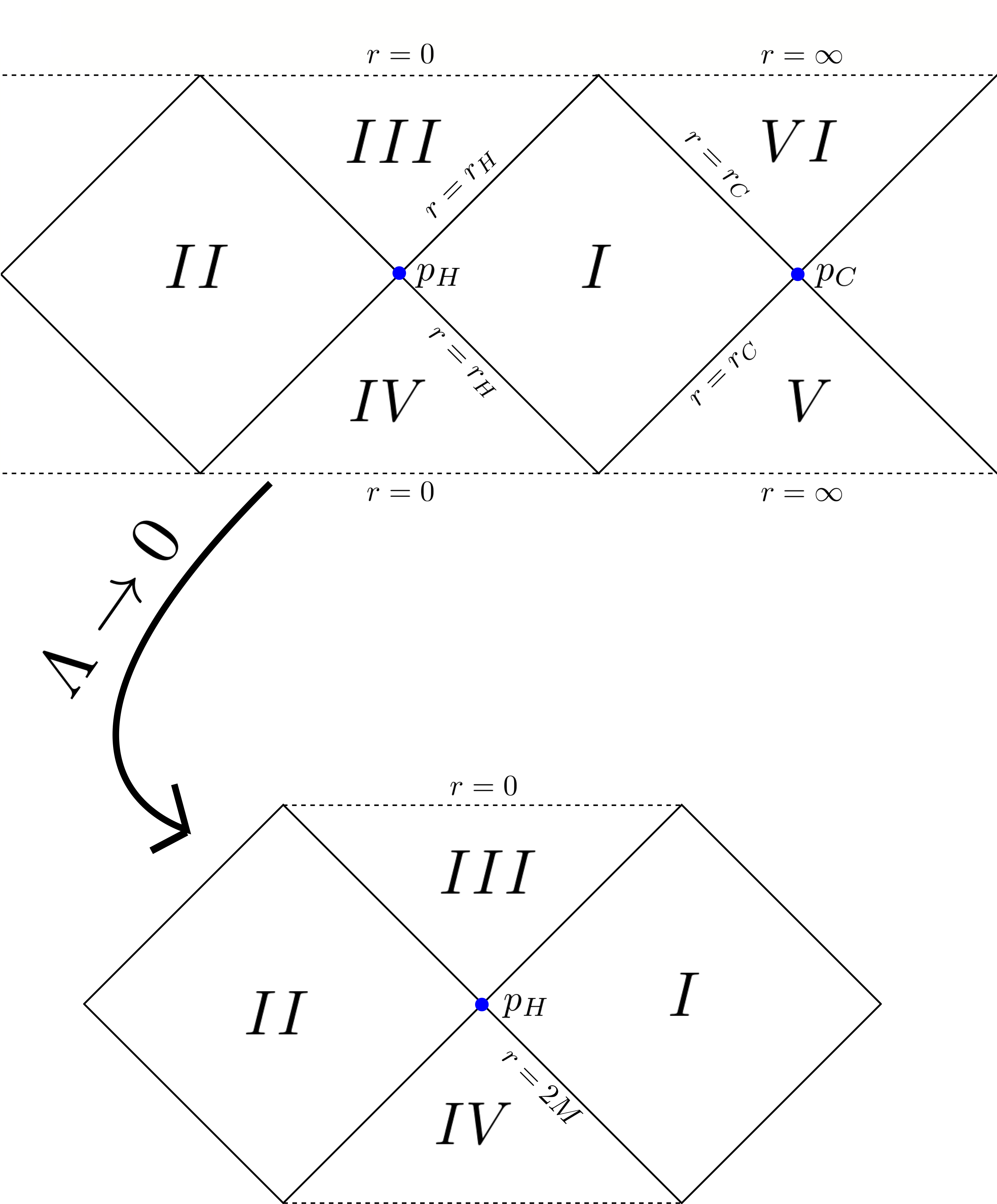}
    \caption{The conformal diagram of Schwarzschild-de Sitter and the conformal diagram of the limiting Schwarzschild spacetime.}
    \label{fig:ssdstoss}
\end{figure}

To illustrate, let us look at the limit of Schwarzschild-de Sitter, as the value of the cosmological constant goes to zero, and take as our fiducial point the point bifurcation sphere $p_H$, shown in Figure \ref{fig:ssdstoss}. A natural question is to ask whether points in block VI exist in the limit. Geodesics from $p_H$ to a point $p_6$ in region VI must first pass through $r = r_C$. That is,
\begin{align*}
d(p_H,p_6) = d(p_H,r_C) + d(r_C,p_6).
\end{align*}
But the first term diverges in the limit, so 
\begin{align*}
\lim_{\Lambda \to 0} d(p_H,p_6) = \infty,
\end{align*}
and it follows that $p_6$ cannot survive in the limit. A similar argument shows that region V cannot survive either.
Note that choosing $p_C$ as the fiducial point could result in a completely different limiting spacetime, however we will not investigate this question in the present work. 

%%%%%%%%%%%%%%%%%%%%%%%%%%%%%%%%%%%%%%%%%%%%%%%%%%%%%%%%%%%%%%%%%%%%%%%%%%%%%%%%%%%%%%%%%%%%%%%%%%%%%%%%%%%%%%%%%%%%%%%%%%

\section{An embedding into anti-de Sitter space}
\label{sec:embedding}

Geroch's notion of limits of spacetime is somewhat abstract, so we shall use the formalism of \cite{bengtsson_classics_2014} to implement the Geroch procedure and describe the associated limits. Following \cite{bengtsson_classics_2014}, we embed the entire one-parameter family of spacetimes into a fixed ambient space, which we take to be $AdS_3$. Each spacetime touches at a definite point in the ambient space, the origin of the $AdS_3$ space, and the tangent spaces (and therefore an othornomal frame) coincide at that point. It follows that the conditions of Geroch's limit theorem are met, and we can therefore uniquely assign a limiting spacetime. Of course, the limiting spacetime will depend on the points we are identifying, that is, on the embedding. There is, in general, no canonical procedure for selecting points in the different spacetimes which we may regard as ``the same''. We will choose this fiducial point to be a point on the bifurcation sphere, $p_H$.

Since our family of spacetimes is spherically symmetric, it is enough to embed the 1+1 dimensional spacetime $\Sigma$, described by the metric
\begin{align}
\label{eq:2DBHmetric}
ds^2 &= - f(r) dt^2 + \frac{1}{f(r)} dr^2.
\end{align}
The embedding of $\Sigma$ into $AdS_3$ is determined by the following equations
\begin{subequations}
\label{eq:AdS3embedding}
\begin{align}
X &= \sqrt{1+a^2 f(r)} \, \sinh{\left( g(r) \right)} \\
Y &= a \sqrt{f(r)} \cosh{\left( \frac{t}{a} \right)} \\
U &= a \sqrt{f(r)} \sinh{\left( \frac{t}{a} \right)} \\
V &= \sqrt{1+a^2 f(r)} \, \cosh{ \left( g(r) \right)}.
\end{align}
\end{subequations}
The parameter $a$ is a constant which we choose for convenience to be $\frac{1}{\kappa}$, where $\kappa$ is the surface gravity of the black hole. The functions $(X,Y,U,V)$ are coordinates for the $AdS_3$ space, thought of as the hypersurface $X^2 +Y^2 - U^2-V^2 = -1$ in $\mathbb{R}^4$, endowed with the metric
\begin{align}
\label{eq:AdS3metric}
ds^2 &= dX^2 + dY^2 - dU^2 - dV^2.
\end{align}
Since we want this embedding to be an isometric embedding of our spacetime into $AdS_3$, we insist that the induced metric, determined by the ambient $AdS_3$ metric (\ref{eq:AdS3metric}) and the embedding (\ref{eq:AdS3embedding}), matches the black hole metric (\ref{eq:2DBHmetric}). This will occur when the function $g(r)$ satisfies the differential equation
\begin{align}
\label{eq:embeddingDE}
\big( g'(r) \big)^2 &= \frac{1 + a^2 f - \frac{a^2 f'}{4}}{f \big( 1+a^2f \big)^2}.
\end{align}
Note that, so far, the only difference between the setup here and the setup in \cite{bengtsson_classics_2014} is the form of the function $f(r)$. Determining the embedding therefore amounts to solving the differential equation (\ref{eq:embeddingDE}) for the function $g(r)$, which we will do numerically. By choosing $g(r_H) = 0$, we are able to ensure that the black hole horizon for each embedding touches the point $(X,Y,U,V) = (0,0,0,1)$ in the ambient $AdS_3$ space. 

In order to visualise the embeddings, we will use the so-called sausage coordinates $(x,y,\tau)$ for $AdS_3$. These coordinates are related to the embedding coordinates $(X,Y,U,V)$ by:
\begin{equation*}
\begin{aligned}[c]
X &= \frac{2x}{1-\rho^2} \\[1em]
Y &= \frac{2y}{1-\rho^2} 
\end{aligned}
\qquad
\begin{aligned}[c]
U &= \frac{1 + \rho^2}{1-\rho^2} \sin \tau \\[1em]
V &= \frac{1 + \rho^2}{1-\rho^2} \cos \tau,
\end{aligned}
\end{equation*}
where $\rho = x^2 + y^2$ and $0\leq \rho < 1$. The sausage coordinates realise $AdS_3$ as a solid cylinder in $\mathbb{R}^3$. Slices of constant $\tau$ in this cylinder are Poincar\'{e} disks, and the embedding of $\Sigma$ into the $AdS_3$ space now appears as a two dimensional sheet inside the solid cylinder. We refer the reader to the appendix of \cite{bengtsson_classics_2014} for a nice discussion of the geometric properties of this embedding. 
\\

%%%%%%%%%%%%%%%%%%%%%%%%%%%%%%%%%%%%%%%%%%%%%%%%%%%%%%%%%%%%%%%%%%%%%%%%%%%%%%%%%%%%%%%%%%%%%%%%%%%%%%%%%%%%%%%%%%%%%%%%%%

\section{Illustrations}
\label{sec:pics}

\begin{figure}
    \centering
    \includegraphics[width=0.8\textwidth]{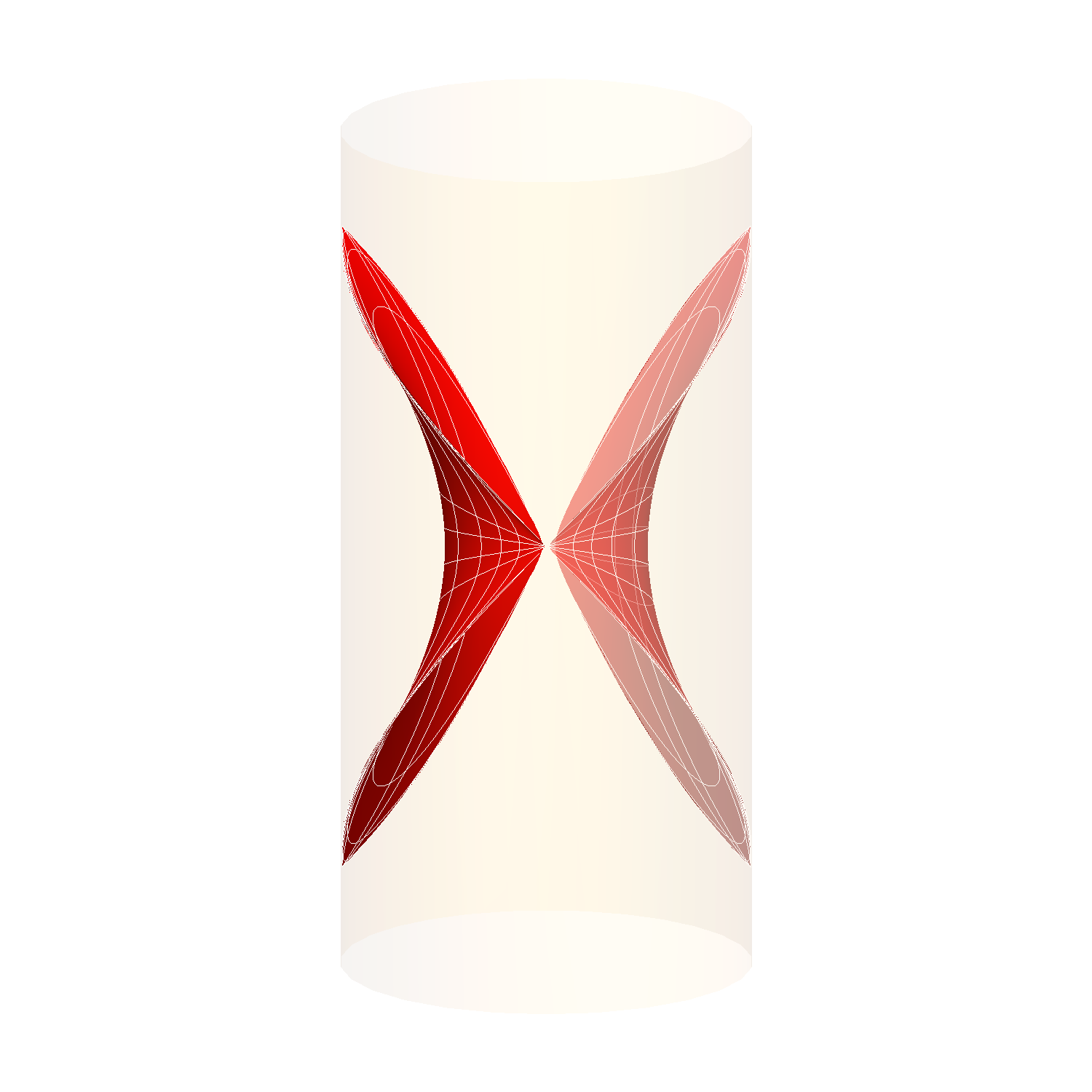}
    \caption{An embedding of Schwarzschild-de Sitter, with $\Lambda M^2 = \frac{1}{10}$. The $AdS$ cylinder in being viewed from the left. One of the sheets has been made translucent to aid visualisation.}
    \label{fig:basicssdsembeddinga}
\end{figure}

\begin{figure}
    \begin{multicols}{2}
    \includegraphics[width=0.5\textwidth]{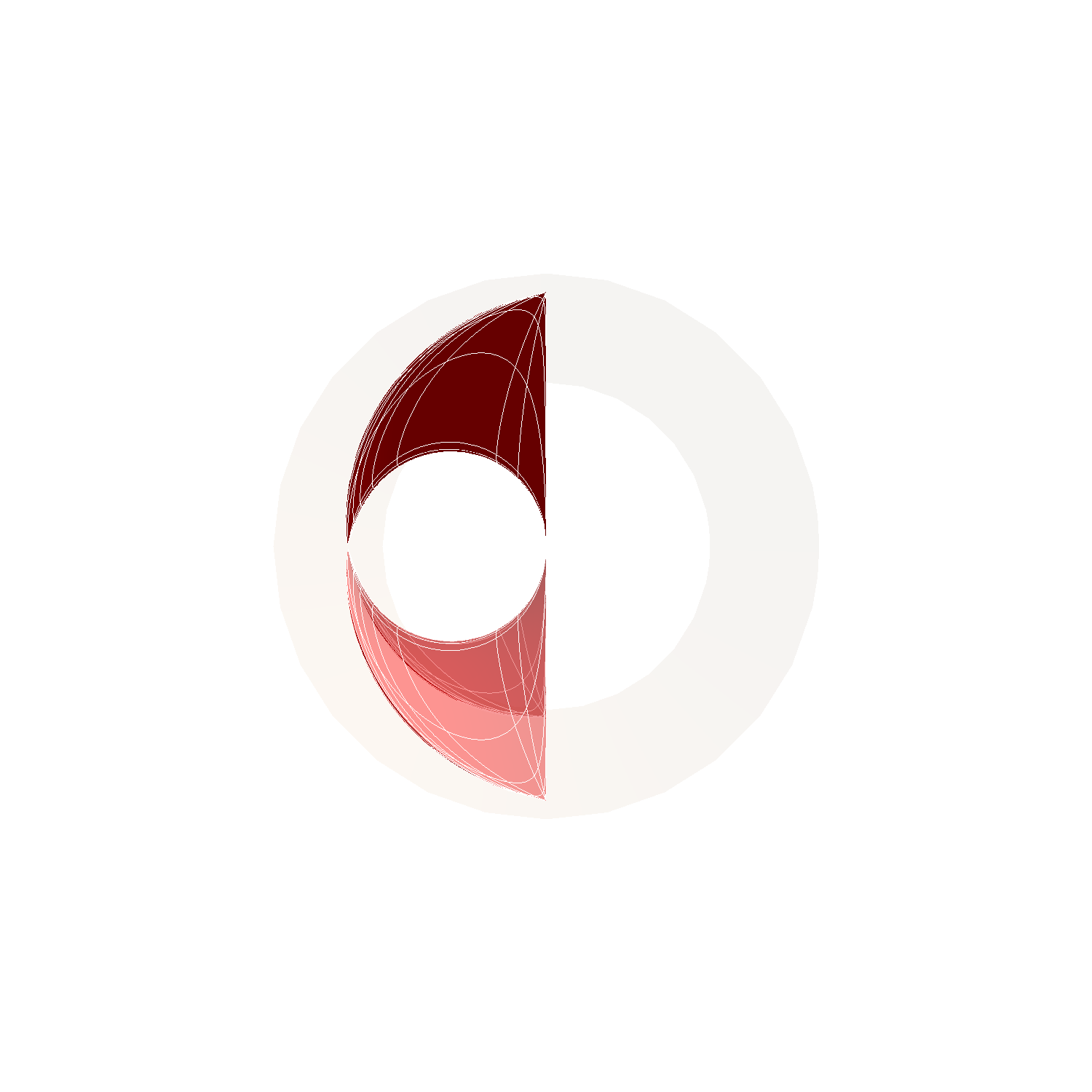}\par
    \includegraphics[width=0.5\textwidth]{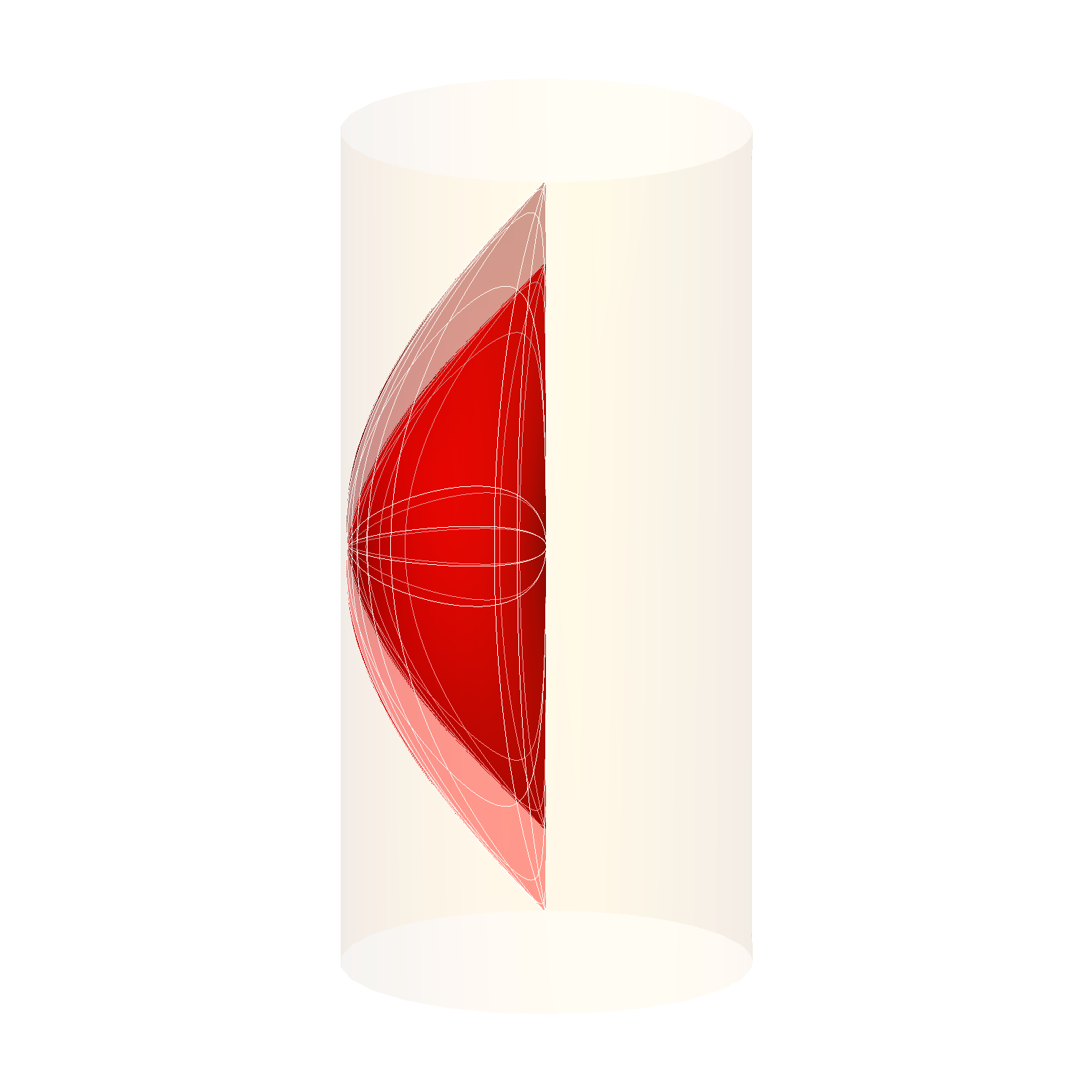}
    \end{multicols}
    \caption{Views of the embedding in Figure \ref{fig:basicssdsembeddinga} from above (image on left) and the front (image on right). Note that these figures have been produced in Mathematica from a three-dimensional figure, and the pictures are stereographic projections from the described viewpoints.}
        \label{fig:basicssdsembeddingb}
\end{figure}

When we embed a Schwarzschild-de Sitter spacetime, we have to choose the values of $M$ and $\Lambda$ for a given embedding. A straightforward way to do this is to fix $M$ to some convenient value, say $M = 1$, and then study the embeddings as you vary $\Lambda$. A representative embedding is shown in Figures \ref{fig:basicssdsembeddinga} and \ref{fig:basicssdsembeddingb}.

\begin{figure}
    \centering
    \includegraphics[width=0.9\textwidth]{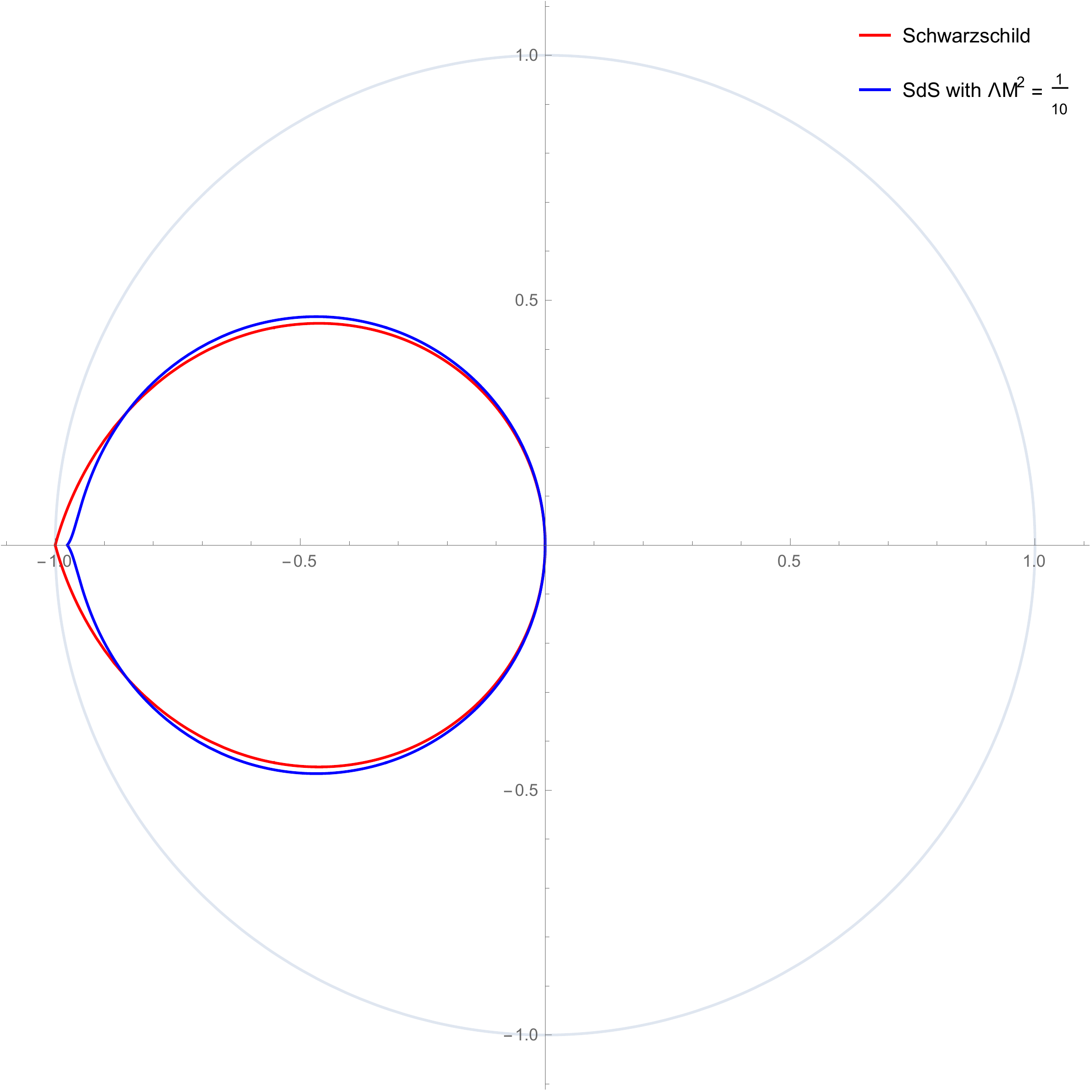}
    \caption{The $\tau = 0$ slice of the embedding in Figure \ref{fig:basicssdsembeddinga}, and the embedding of Schwarzschild of the same mass. The two embeddings touch at the origin of the ambient $AdS_3$ space.}
    \label{fig:basiccircles}
\end{figure}

In Figure \ref{fig:basiccircles}, we plot the $\tau = 0$ slice of this embedding, together with the $\tau = 0$ slice of the Schwarzschild embedding of the same mass. 

\begin{figure}
    \centering
    \includegraphics[width=0.9\textwidth]{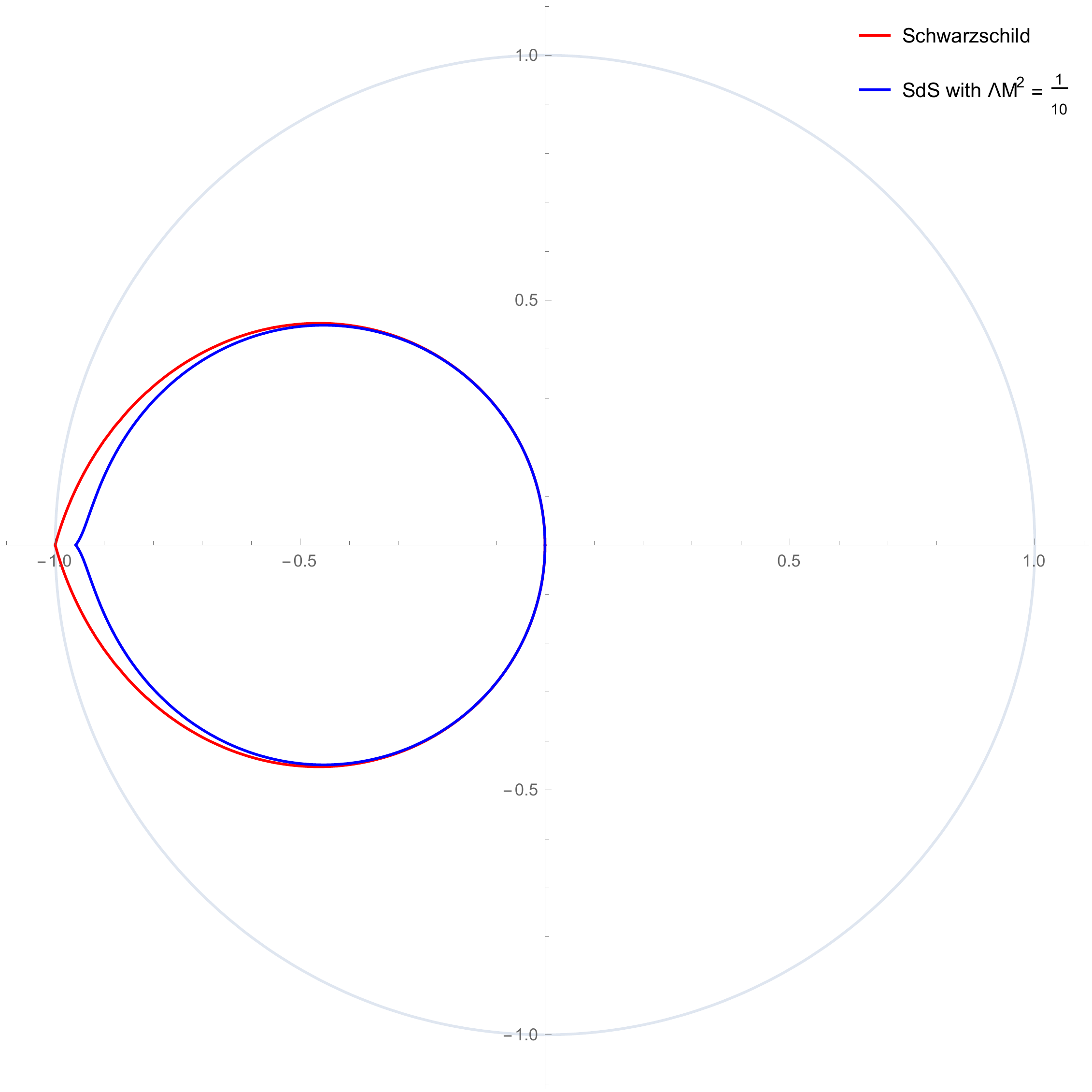}
    \caption{The $\tau = 0$ slice of the embedding in Figure \ref{fig:basicssdsembeddinga}, and the embedding of the mass-corrected Schwarzschild. The two embeddings touch at the origin of the ambient $AdS_3$ space.}
    \label{fig:masscorrectedcircles}
\end{figure}

An unpleasant feature of this picture is the discrepancy between the embedding of the Schwarzschild-de Sitter domain of outer communication and the embedding of the Schwarzschild domain of outer communication. The physical interpretation discussed in Section \ref{sec:phys} involves a comparison between the near horizon geometry of  Schwarzschild and Schwarzschild-de Sitter black holes. The key point is that when comparing these black holes, the near-horizon geometry only matches once we adjust the relative masses. To achieve this, we consider a mass parameter $M = M(\Lambda)$, varying with $\Lambda$ in such a way that the horizon area is kept constant. That is, we want to fix the radius of the black hole horizon to be $r = r_H = 2 \mu$, where $\mu$ is the mass of some reference Schwarzschild spacetime. Note that this mass-fixing procedure is equivalent to changing the mass of the reference Schwarzschild black hole. The $\tau = 0$ slice of the resultant embeddings provide a much cleaner comparison between the Schwarzschild and the Schwarzschild-de Sitter embeddings, as seen in Figure \ref{fig:masscorrectedcircles}. We shall employ this mass-correction for the remainder of the present work. 

Before we elaborate more on the physical interpretation, let us first make a few comments on how to view the embeddings we have already obtained. The two sheets of the embedding in Figures \ref{fig:basicssdsembeddinga} and \ref{fig:basicssdsembeddingb} correspond to regions I and II in the Schwarzschild-de Sitter conformal diagram in Figure \ref{fig:ssds}. In the $\tau = 0$ slice, the centre of the disk corresponds to the origin of the $AdS_3$ space, and the event horizon of the embeddings. The circle $x^2 + y^2 = 1$, corresponding to the boundary of the solid cylinder in Figures \ref{fig:basicssdsembeddinga} and \ref{fig:basicssdsembeddingb}, is an infinite metric distance away from the origin. The blue line is the intersection of the embedding of the Schwarzschild-de Sitter spacetime with this plane, and the other intersection of the blue line with the $y=0$ line corresponds to the cosmological horizon, $r_C$. The fact that the Schwarzschild de Sitter spacetime is a smooth manifold of topology $S^1\times S^2 \times \Reals$ and the embedding is isometric, suggests that the cuspy nature of this intersection is a numerical artefact.  The red line is the intersection of the embedding of the Schwarzschild black hole with the plane $\tau = 0$. Note that the Schwarzschild spacetime reaches the edge of the $AdS_3$ space - points in the Schwarzschild domain of outer communication can be arbitrarily far from the event horizon. 

\begin{figure}
    \centering
    \includegraphics[width=0.9\textwidth]{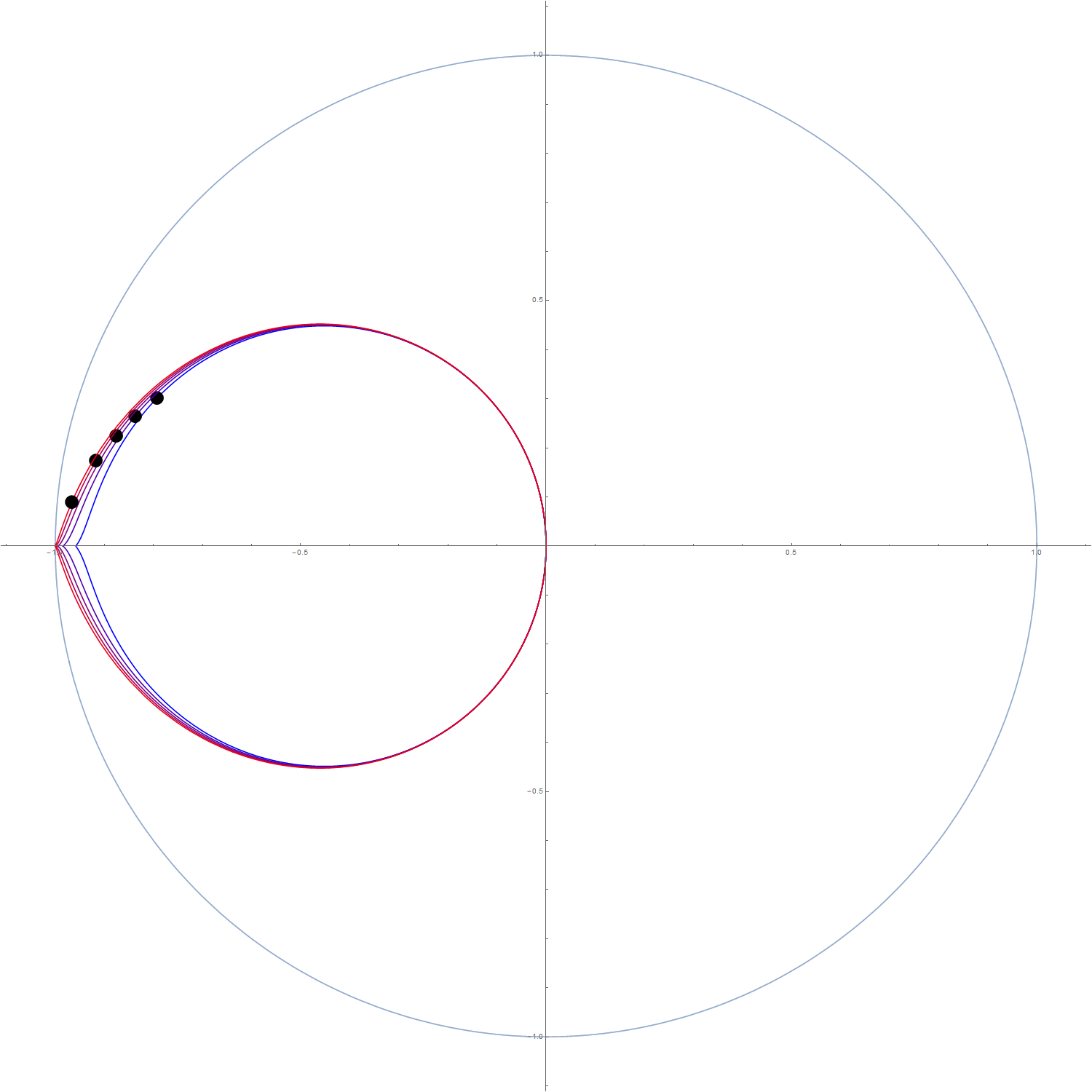}
    \caption{A plot of the $\tau = 0$ slices for various embeddings. The values of $\Lambda$ are such that $9 \Lambda M^2$ is given by $\left[ \frac{9}{10} (\textrm{blue}), \frac{7}{10}, \frac{5}{10}, \frac{3}{10} ,\frac{1}{10} (\textrm{red}) \right] $. The point at which $f'(r) = 0$ is represented on each embedding by a solid dot (See Section \ref{subsec:hierarchy} for more details).}
    \label{fig:differentlambda}
\end{figure}

%%%%%%%%%%%%%%%%%%%%%%%%%%%%%%%%%%%%%%%%%%%%%%%%%%%%%%%%%%%%%%%%%%%%%%%%%%%%%%%%%%%%%%%%%%%%%%%%%%%%%%%%%%%%%%%%%%%%%%%%%%

\section{Physical interpretation}
\label{sec:phys}
In the following section, we will use the illustrations of the previous section to establish a heuristic argument in favour of a hierarchy of validity between the Einstein-de Sitter equations and the Einstein equations. 
\subsection{Schwarzschild mass correction in a de-Sitter Universe}

When embedding the Schwarzschild-de Sitter black holes into $AdS_3$, we had to choose the mass parameter of the black hole to be a function of $\Lambda$ to guarantee that the black hole horizon area remained constant. By identifying the radius of the black hole horizon $r = r_H$ with the radius of a reference Schwarzschild black hole horizon $r = 2 \mu$, we obtain a relation between the mass parameter of the Schwarzschild-de Sitter spacetime $M$ and the effective mass of the reference Schwarzschild black hole $\mu$. Doing this na\"{i}vely by using the expression (\ref{eq:rH}) for $r_H$, we obtain
\begin{equation}
\label{Mlambda}
    M =\frac{1}{3\sqrt{\Lambda}}\cos \Big( 3 \arccos(\mu \sqrt{\Lambda})-\pi \Big)
\end{equation}
Note that since $M = M(\Lambda,\mu)$, the extremality condition $9M \Lambda^2 < 1$ changes, and now becomes 
\begin{align*}
\Lambda < \frac{1}{4 \mu^2}.
\end{align*}
A much simpler expression can be obtained by noting that $f(r_H) = 0$, and so fixing the horizon at $r_H = 2 \mu$ means that we require 
\begin{equation}
f(2\mu) = 1 - \frac{2M}{2\mu} - \frac{\Lambda}{3}\left(2\mu \right)^2 = 0
\end{equation}
Rearranging this expression for $M$ gives us
\begin{equation}
\label{eq:correctedmass}
M = \mu - \frac{4 \Lambda}{3} \mu^3,
\end{equation}
which is identical to the expression (\ref{Mlambda}).
Until this point, we have been using natural units to simplify calculations and expressions. We find it prudent to now switch to S.I. units (meters, kilograms, seconds). Expression (\ref{eq:correctedmass}) for the corrected-mass is, in S.I. units, given by
\begin{equation}
M = \mu - \frac{4 \Lambda G^2}{3c^4} \mu^3.
\end{equation}
For a system with a fixed Schwarzschild/Newtonian mass $\mu$, the Schwarzschild-de Sitter solution with corrected mass $M$ exhibits a similar near field behaviour.

\subsection{Hierarchy of Validity}
\label{subsec:hierarchy}
In quantum mechanics the limit $\hbar \rightarrow 0$ serves to recover the equations governing the evolution of systems in classical mechanics from the equations that govern the same system in the quantum regime. This gives us two things: 
\begin{itemize}
\item A compatibility of quantum mechanics and classical mechanics
\item A breakdown criterion for regimes in which classical mechanics is no longer valid.
\end{itemize}
These two things emphasise that the modeling of a system is scale dependent. Newtonian Gravity emerges from Einstein's Relativity in a similar fashion, namely as a static, small perturbation to a flat background spacetime. We will argue that the $\Lambda \to 0$ limit related the Einstein-de Sitter equations and the Einstein equations in a similar fashio. We will be able to establish a heuristic hierarchy of validity between these systems describing gravity. The precise form in which the embedding of the Schwarzschild-de-Sitter spacetimes approach the asymptotically flat limit further serves to clarify the effect of a non-zero $\Lambda$.

We see from the illustrations in Section \ref{sec:pics} that a non-zero $\Lambda$ mainly effects the structure of the exterior region in the neighbourhood of infinity/the cosmological horizon, that is, regions far away from the massive body. Speaking in interaction terms a non-zero $\Lambda$ effects only the long-range interaction between a massive gravitating object and a test particle. 

Let us now introduce the notion of a radius of validity - namely a radius outside of which the Schwarzschild-de Sitter solution starts to significantly differ from the Schwarzschild solution. We can identify a candidate for such a radius by investigating properties of the radial function $f(r)$. Outside the event horizon, the radial function for Schwarzschild-de Sitter agrees closely with the radial function for Schwarzschild. The point at which they begin to significantly differ is the maximum of the Schwarzschild-de Sitter radial function - that is, the point at which $f'(r) = 0$. This is
\begin{equation}\label{eq:rval}
    r_{v}=\left(\frac{3GM}{c^2 \Lambda}\right)^{\frac{1}{3}}.
\end{equation}
These radii are shown in Figure \ref{fig:differentlambda} for various values of $\Lambda$. Their exact location in the embedding suggests that this is a sensible choice for a radius of validity.
It would be more satisfying to have a geometric characterisation of this radius - that is, a definition that was coordinate independent. We can obtain this by noting that at this radius, we have 
\begin{align*}
\mathcal{R}^2 = 3 \mathcal{I}_1,
\end{align*}
where $\mathcal{R} = 4 \Lambda$ is the Ricci scalar curvature of the Schwarzschild-de Sitter metric, and $\mathcal{I}_1$ is a principal invariant of the Weyl scalar, $C_{abcd}$, defined by
\begin{align*}
\mathcal{I}_1 &= C_{abcd} C^{abcd}.
\end{align*}

\subsection{Comparison of validity length scale with size of astronomical objects}
Since we have obtained a formula for the radius of validity of the Einstein equations in a de Sitter universe, let us now compare that radius of validity to the size of astrophysical objects. As it mostly boils down to an order of magnitude comparison, we have chosen to compare four systems that are roughly representative for their class and cover the various mass ranges. The Solar system, the Globular Cluster NGC 2419, the Milky Way and the Virgo Super Cluster have masses of the order of $1$, $10^5$, $10^{11}$, $10^{15}$ solar masses. Note that we will abstain from using any sort of astronomical units and will be working with SI units instead.

\begin{table}[h]
\begin{center} 
 \begin{tabular}{|c | c | c | c|} 
 \hline  
 Object  & Mass (kg) & Size (m) & $r_v$ (m) \\ 
 \hline 
 Solar System  & $2\times10^{30}$ & $7.5\times10^{12}$ & $3\times 10^{18}$ \\ 
 \hline
 NGC2419 Globular Cluster & $2\times10^{36}$ & $2.5\times 10^{18}$  & $3\times 10^{20}$  \\
 \hline
 Milky Way (with out dark matter halo) & $2\times 10^{41}$ & $9.5\times10^{20}$  & $1.5 \times 10^{22}$\\
 \hline
  Milky Way (with dark matter halo) & $2\times 10^{42}$ & $1.5\times10^{21}$  & $3 \times 10^{22}$\\
 \hline
 Virgo Supercluster & $2\times10^{45}$ & $5.2\times10^{23}$  & $3 \times 10^{23}$ \\
 \hline
 Universe (present day) & $3 \times 10^{52}$ & $4.3\times10^{26}$  & $8 \times 10^{25}$ \\  
 \hline
\end{tabular}
\end{center}
\caption{Observational values of astronomical systems compared to the scale of validity calculated by formula \eqref{eq:rval}. A more extensive discussion is contained in the bulk of the text.}
\label{tab:rval}
\end{table}

\begin{figure}
    \centering
    \includegraphics[width=0.8\textwidth]{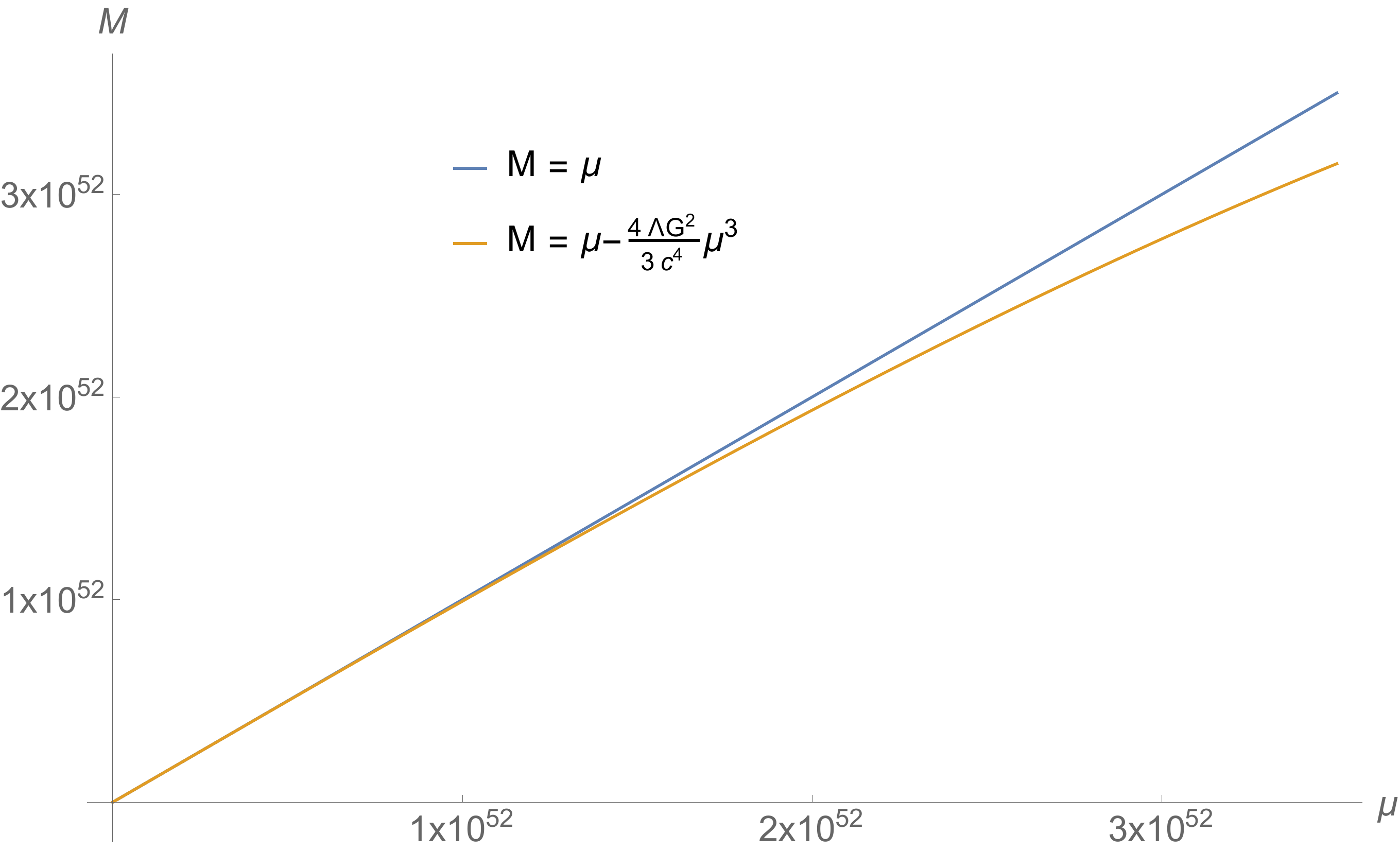}
    \caption{A plot of the Schwarzschild mass $M = \mu$ and a plot of the corrected mass, as determined by (\ref{eq:correctedmass}).}
    \label{fig:masscorrection}
\end{figure}

First we observe that the solar system and the globular cluster both have radii of validity that extend well beyond their physical size. Secondly, since the mass of the system affects the radius of validity, we calculated the radius of validity for the Milky Way with and without dark matter. Dark matter was originally introduced to compensate for deviations from a simple Newtonian calculation. Now since the Newtonian approximation is an approximation to the Einstein field equations, its application beyond the radius of validity for the Einstein field equation is delicate. The radius of validity depends on the total mass of the system, so if one adds in dark matter to fix deviations from Newtonian calculations, one artificially extends the radius of validity. If the radius of validity for a system without dark matter were smaller than the size of the system, such an artificial extension might cause one to wrongfully conclude that the system lies within the radius of validity.\\
Assuming for the sake of simplicity that dark matter makes up 90\% of the total mass of the Milky Way, it changes the radius of validity roughly by a factor of $2$. Given that the proposed dark matter halo of the Milky Way also extends significantly further out then just the edge of the disk, the ratio between the systems size and the radius of validity barely changes. However in both cases the radius of validity is roughly one order of magnitude bigger than the physical size of the system and thus we conclude that using the Einstein field equations and thus Newtonian calculations is adequate.\\
For the Virgo Super Cluster, however, the radius of validity is of the same order of magnitude as the system. In fact, the radius of validity is roughly half the radius of the system itself. This implies that applying Newtonian or post-Newtonian calculations to that system has to be done with care. The fact that we are using here a point particle and spherically symmetric approach for a system as extended as the Virgo Super Cluster means that we can not make strict statements on whether such calculations are actually invalid or not.\\
While the point particle approach for the Virgo Super Cluster is still somewhat justified, the same can not be said about the Universe as a whole. There we see that for the present day universe the radius of validity is an order of magnitude smaller than its size. In that case instead of using equation \eqref{eq:rval} for a given mass $M$, we replace $M$ by the mass contained in a sphere of Radius $\textbf{R}$ with homogeneous matter density $\rho$.\footnote{In an abuse of notation, we will in the following compare a sphere of radius $\textbf{R}$ in Euclidean space with a coordinate sphere in Schwarzschild-de Sitter.} That is, we replace $M$ in \eqref{eq:rval} with 
\begin{align}
M = \frac{4 \pi}{3}\textbf{R}^3 \rho  .  
\end{align}
Rearranging \eqref{eq:rval} we then obtain 
\begin{equation}\label{eq:critdens}
    \frac{r_v}{\textbf{R}}= \left(\frac{4\pi G \rho}{c^2 \Lambda}\right)^{1/3} 
\end{equation}
which is bigger than $1$ whenever $\frac{4\pi G \rho}{c^2 \Lambda}>1$. In cosmology the different eras (radiation-dominated, matter-dominated, $\Lambda$-dominated) are distinguished by the type of energy (radiation, matter, $\Lambda$) that makes up the largest fraction of the total energy. We see then, that (\ref{eq:critdens}) tells us that the radius of validity for a system outside the radius of the system, and therefore the Einstein field equations are a valid approximation, precisely when it is matter-dominated. Hence we could arrive at a similar expression for the redius of validity by looking at the ratio $\rho_M/\rho_{vac} $  between the matter density $\rho_M=\frac{3M}{4\pi r^3}$ of a mass $M$ evenly distributed over a sphere of radius $r$, and the vacuum energy density $\rho_{vac}=\frac{\Lambda c^2}{8 \pi G}$ we get
\begin{equation}
    \frac{\rho_M}{\rho_{vac}}= \left(\frac{6 M G }{\Lambda c^2 r^3}\right).
\end{equation}
This is equal to $1$ precisely when 
\begin{equation}
    r=2^{1/3}r_v.
\end{equation}
Thus for a given mass $M$, the the radius of validity is, up to a small numerical factor, the same radius at which the matter density and vacuum energy density are equal. For $r<r_v$, the matter density dominates and we are confident that the Einstein equations provide a good description. For $r>r_v$, the vacuum energy dominates, and we should be careful about applying Newtonian or post-Newtonian arguments. \\
Up to this point we have ignored the mass-correction formula because for all the compact systems under consideration so far it  was sufficient to take the approximation $M = \mu$. It is only on mass scales that are on the order of the mass of the universe that we see a significant deviation, as can be seen in Figure \ref{fig:masscorrection}. Indeed, a mass deviation of $1$\% only occurs once the mass of the system reaches $10^{52}$ kg. For $\mu = 3 \times 10^{52}$, which is roughly the mass of the observable universe, the mass-correction is roughly 10\%. This can be thought of as a secondary modification that takes effect already within the radius of validity and might thus be relevant for considerations on the scale of the universe. Here of course one has to keep in mind that the mass-correction formula originates from a point particle consideration and is thus not necessarily applicable to the universe. Note also that the mass-correction becomes negligible when we only consider baryonic matter. On the other hand in the early universe, when the total energy from the electromagnetic radiation was significantly higher the effect might be more prominent. 

\section{Conclusion}
We derived an isometric embedding for the Schwarzschild and Schwarzschild-de Sitter spacetimes into $AdS_3$. We used the detailed behaviour of the embedding in the $\Lambda \to 0$ limit to heuristically define a radius of validity for the Einstein Equations in a de Sitter universe. One possible interpretation of this hierarchy of validity is that one can assign to $\Lambda$ a similar role in the context of gravity that $\hbar$ plays for quantum mechanics. This observation suggests that one can, in principle, interpret the cosmological constant as a fundamental energy scale for gravitational systems.\\
The considerations in Section \ref{sec:phys} show that for most scales in the universe, it is safe to ignore possible effects of the cosmological constant. For large systems, however, using an Einstein or Newtonian approximation may not be justified, despite the low value of the cosmological constant. In particular for the largest structures such as superclusters, the Newtonian approximation might not be entirely valid. Note that these effects on long-range interactions could affect the interpretation of weak lensing observations, since most of the reconstruction is based on post Newtonian approximations, see for example \cite{kilbinger2015cosmology} for an extensive review. Note in particular that beyond the radius of validity the sign of $f'(r)$ changes and thus the lensing might behave substantially different beyond this point. Similar considerations for the long range interactions might be relevant to figure out which black hole stability problem is actually the physical relevant one. (i.e. Kerr or Kerr-de-Sitter). Last but not least, the field of vision for LIGO spans far beyond the radius of validity of even the Virgo Supercluster \cite{abbott2016prospects}. Modifications of gravitational wave sources in the spirit of \cite{ashtekar2014asymptotics,ashtekar2015asymptotics,ashtekar2015asymptotics2,ashtekar2016gravitational} might therefore have an effect on observations.  \\
For a homogeneous universe the cosmological constant becomes relevant when the matter density and the vacuum energy density are roughly equal. However there is a secondary effect due to the mass-correction that might play a role. However it is at most of the order of 10\% so it is certainly no dramatic change. 

One limitation to the present work is, that \emph{a priori} it only holds true for the case of spherical symmetry which we investigated here. This is relevant to mention because preliminary calculations for an extension of the results in  \cite{mars2017fingerprints} to the case with a positive cosmological constant suggest that in principle $\Lambda$ is detectable in the shape of the shadow of a black hole when $a>0$. As the shadow contains mostly near horizon information, this suggests that the cosmological constant should affect the near horizon geometry. 

It would be interesting to try to elaborate in a quantitative manner on the features investigated in the present work. Further it would be interesting to investigate the role of the cosmological constant away from spherical symmetry.
%%%%%%%%%%%%%%%%%%%%%%%%%%%%%%%%%%%%%%%%%%%%%%%%%%%%%%%%%%%%%%%%%%%%%%%%%%%%%%%%%%%%%%%%%%%%%%%%%%%%%%%%%%%%%%%%%%%%%%%%%%

\subsection*{Acknowledgements}This work was partially supported by the Australian Research Council grant DP170100630.
We would like to thank Hermine Boat for her essential support during the conceptional phase of this paper. We would further like to thank the Institute Henri Poincar\'{e} for hospitality during the trimester on Mathematical Relativity, Paris, during fall 2015, and M.B. would like to thank Monash University where part of this work was done. C.P. was supported by the Albert Einstein Institute during a part of this project. M.B. was supported by an Australian Postgraduate Award for part of this project. 
\newpage

\bibliographystyle{plain} %plain %plainnat
\bibliography{spacetimelimits.bib}

\end{document}